\documentclass[twocolumn,aps,superscriptaddress,pre]{revtex4}

\usepackage{amsmath,amssymb,graphicx}
\usepackage{subfigure}
\usepackage{algorithmic,algorithm} 
\usepackage{enumerate}
\usepackage{times}
\usepackage{color}
\usepackage{soul}
\definecolor{yblue}{rgb}{0.06, 0.3, 0.57}
\usepackage[pdftex]{hyperref}
\hypersetup{colorlinks=true,linkcolor=blue,citecolor=blue,urlcolor=blue}

\newcommand{\mean}[1]{\langle{#1}\rangle}

\begin{document}

\title{An effective introduction to the Markov Chain Monte Carlo method}

\author{Wenlong Wang}
\email{wenlongcmp@scu.edu.cn}
\affiliation{College of Physics, Sichuan University, Chengdu 610065, China}

\begin{abstract}
We present an intuitive, conceptual, but semi-rigorous introduction to the celebrated Markov Chain Monte Carlo method using a simple model of population dynamics as our motivation and focusing on a few elementary distributions. Conceptually, the population flow between cities closely resembles the random walk of a single walker in a state space. We start from two states, then three states, and finally the setup is fully generalized to many states of both discrete and continuous distributions. Despite the mathematical simplicity, the setup remarkably includes all the essential concepts of Markov Chain Monte Carlo without loss of generality, e.g., ergodicity, global balance and detailed balance, proposal or selection probability, acceptance probability, up to the underlying stochastic matrix, and error analysis. Our teaching experience suggests that most senior undergraduate students in physics can closely follow these materials without much difficulty.
\end{abstract}

\maketitle

\section{Introduction}
Computational physics plays an increasingly important role in physics research, and it is frequently referred to as the third branch of physics in addition to theoretical physics and experimental physics \cite{Binder:MC,UPA}. This is largely because of the enormous progress made in the past decades in both the growth of computational power and the development of numerical algorithms, as well as the flexibility of computational physics. Numerical algorithms come in various classes, depending on the nature of the underlying mathematical problem. Monte Carlo (MC) method is a large set of algorithms for statistical sampling a given distribution, it can solve virtually any problem as long as one can map it onto a probability one. It is particularly relevant in statistical physics, but it is also widely used in solving optimization problems and running various stochastic dynamics \cite{SA,Wang:GS,Glauber:MC,DLA}. Because the MC method is generic, it is also used in many other fields such as statistics and biology.

The central idea of the Monte Carlo method is the Markov Chain Monte Carlo (MCMC) \cite{metropolis:49,metropolis:53}. A state or a walker does weighted random walk in a state space, generating a chain or a time series of states. MCMC has various extensions, e.g., in the extended-ensemble methods for sampling glassy systems, several related Markov chains at different parameters can be coupled together \cite{Hukushima:PT}. MCMC is also central to the so-called sequential Monte Carlo \cite{SA,Hukushima:PA,Machta:PA,Wang:PA,Weigel:PA}.

There exist many excellent introductory materials on the MCMC at various levels \cite{AA:comp,Mark:comp,Walter:MC,newman:99,Binder:MC,NP:MC,Liu2004}. Our work is clearly not a repetition of any of these materials. Our motivation is that many such introductory materials in physics are quite fast in pace for undergraduate students, while others may only briefly mention it. This paper aims to fill this important gap as much as possible for a smoother transition using simple examples. Indeed, it is a quite common practice that after introducing the idea and the basic rules of the MCMC method, the students are posted with the famous Ising model \cite{Ising:25,brush:67} or classical particle systems \cite{ca:patterns,Wang:VC}, and so on. While this approach clearly has the advantage that more advanced topics can be covered, it also has a complementary drawback that a student may not be able to fully digest the principles of the MCMC method. For example, sometimes the underlying stochastic matrix behind the MCMC is entirely omitted and the understanding is fully based on intuition. In addition, this approach needs to deal with a number of technical aspects that are not really related to the Monte Carlo principles and are also not necessarily familiar to students, e.g., what is the Ising model, and what is a periodic boundary condition?
Therefore, while this approach could be utilized for students who already have a suitable background, it is in our opinion not ideal for students who are studying MCMC for the first time, we find that it is easier if they are introduced later. It is noted that there are also many introductory articles on MCMC in other fields, e.g., statistics \cite{SMC}, but the models and applications therein are frequently not very familiar to physicists.

The main purpose of this article is to present an intuitive, conceptual, but semi-rigorous introduction to the Markov Chain Monte Carlo method using simple distributions such that it is accessible to undergraduate students in science particularly in physics. Indeed, we start from two states in detail, explaining the existence of an equilibrium of a random walk dynamics and then reverse the idea for statistical sampling. Next, we study three states, introducing more concepts such as ergodicity, and balance conditions. Here, a good number of examples are given to practise the design of a random walk. The underlying stochastic matrix of the MCMC is then discussed. The setup is subsequently generalized to many states for the discrete geometric distribution and the continuous Gaussian distribution. Finally, the data analysis of correlated data is presented. It is worth mentioning that we are not proposing these elementary distributions merely because they are simple, the beauty of this approach is that we can discuss essentially all the pertinent concepts of MCMC in our framework. It seems likely that the materials presented herein are understandable to many undergraduate students also in related fields such as mathematics, and chemistry.


This work assumes that a reader is already familiar with the simple sampling of uniform distributions, and its limitations. For example, we expect that a student is comfortable with random numbers, the uniform distribution $U[0,1]$ and distribution transformations, and the central limit theorem (CLT) for analyzing \textit{iid} (independent and identically distributed) data \cite{Statinference}. While the MCMC data are correlated, we can still apply the CLT in indirect ways.

\section{Two states}
We start from the simplest Bernoulli distribution of only two states \cite{Statinference} and examine in detail an example of population dynamics. This example illustrates the essential idea of the MCMC method. Consider two cities A and B, and the people living therein decide independently each year whether they stay in the same city or move to the other city with certain probabilities. Suppose that the city A is overall more attractive, but the city B still has its own advantages. If a person in the city A moves to the city B with a probability $p=0.1$ and therefore remains in the city A with a probability $1-p=0.9$, and a person in the city B moves to the city A with a probability $q=0.8$ and consequently stays in the city B with a probability $1-q=0.2$. What is the fate of this dynamics? Interestingly, the answer is that the population will remarkably reach an equilibrium in the long run from any initial condition.

Let us look at the mathematics of this dynamics. The population distribution in year $n+1$ depends on the population distribution in the year $n$ as:
\begin{align}
    \begin{pmatrix}
N_{An+1} \\ 
N_{Bn+1}
\end{pmatrix}
=
\begin{pmatrix}
0.9 &0.8 \\
0.1 &0.2
\end{pmatrix}
\begin{pmatrix}
N_{An} \\ 
N_{Bn}
\end{pmatrix}.
\end{align}
If we only focus on the relative population size, we can normalize them as probabilities:
\begin{align}
    \begin{pmatrix}
p_{A} \\ 
p_{B}
\end{pmatrix}
&=
\frac{1}{N_A+N_B}
\begin{pmatrix}
N_A \\ 
N_B
\end{pmatrix}, \\
\begin{pmatrix}
p_{An+1} \\ 
p_{Bn+1}
\end{pmatrix}
&=
\begin{pmatrix}
0.9 &0.8 \\
0.1 &0.2
\end{pmatrix}
\begin{pmatrix}
p_{An} \\ 
p_{Bn}
\end{pmatrix}.
\label{iterationTS}
\end{align}
The question becomes what is the fate of this iteration?
The long term dynamics is as:
\begin{align}
    \begin{pmatrix}
p_{An} \\ 
p_{Bn}
\end{pmatrix}
=
\begin{pmatrix}
0.9 &0.8 \\
0.1 &0.2
\end{pmatrix}
\begin{pmatrix}
p_{A_{n-1}} \\ 
p_{B_{n-1}}
\end{pmatrix}
=
\begin{pmatrix}
0.9 &0.8 \\
0.1 &0.2
\end{pmatrix}^n
\begin{pmatrix}
p_{A_{0}} \\ 
p_{B_{0}}
\end{pmatrix}.
\end{align}
It is clearly a good idea to find the eigenvalues and eigenvectors of the matrix. The eigvenvalue problem of this $2\times2$ matrix $S$ can be straightforwardly solved:
\begin{align}
\begin{pmatrix}
0.9 &0.8 \\
0.1 &0.2
\end{pmatrix}
\begin{pmatrix}
\alpha \\ 
\beta
\end{pmatrix}
=
\lambda
\begin{pmatrix}
\alpha \\ 
\beta
\end{pmatrix}, \\
    \lambda_1=1, \ 
v_1=
\begin{pmatrix}
0.8 \\ 
0.1
\end{pmatrix}, \\
\lambda_2=0.1, \ 
v_2=
\begin{pmatrix}
1 \\ 
-1
\end{pmatrix}.
\end{align}
The two eigenvectors are linearly independent, so any valid initial vector can be written as a linear combination of the two eigenvectors:
\begin{align}
    \vec{p}_n &=S^n \vec{p}_0 = S^n (c_1 v_1 + c_2 v_2), \\
     &= c_1 S^nv_1 + c_2 S^nv_2 = c_1 \lambda_1^nv_1 + c_2 \lambda_2^nv_2.
\end{align}
Note that $\lambda_1=1$, it does not matter if we repeatedly multiply its eigenvector with $S$. The second part, however, has $|\lambda_2|=0.1<1$, if we keep the iteration, this term will be exponentially suppressed in $n$ and converge to $0$. Therefore, the vector will eventually converge to the \textbf{eigenvector of the largest eigenvalue} from any valid initial condition. Note that we cannot start purely from $v_2$, it has negative components, and it is therefore not a valid initial condition. The population distribution will eventually settle to the equilibrium $(p_A^*, p_B^*)=(8/9, 1/9)$. A few typical examples are illustrated in Table~\ref{para}.

\begin{table}[htb]
\caption{
Convergence to the equilibrium state $(p_A^*,p_B^*)=(8/9,1/9)$ from any initial condition for the two-state system of Eq.~\eqref{iterationTS}. Here, $n$ is the iteration step.
\label{state2}
}
\begin{tabular*}{\columnwidth}{@{\extracolsep{\fill}} l c c c}
\hline
\hline
$n$	&Run 1	&Run 2	&Run 3 \\ 
\hline
$0$	&$(1.0000, 0.0000)$	&$(0.0000, 1.0000)$	&$(0.2500, 0.7500)$ \\ 
\hline 
$1$	&$(0.9000, 0.1000)$	&$(0.8000, 0.2000)$	&$(0.8250, 0.1750)$ \\ 
\hline
$2$	&$(0.8900, 0.1100)$	&$(0.8800, 0.1200)$	&$(0.8825, 0.1175)$ \\ 
\hline
$3$	&$(0.8890, 0.1110)$	&$(0.8880, 0.1120)$	&$(0.8883, 0.1117)$ \\ 
\hline
$4$	&$(0.8889, 0.1111)$ &$(0.8888, 0.1112)$	&$(0.8888, 0.1112)$ \\ 
\hline
$5$	&$(0.8889, 0.1111)$	&$(0.8889, 0.1111)$	&$(0.8889, 0.1111)$ \\ 
\hline
$6$	&$(0.8889, 0.1111)$	&$(0.8889, 0.1111)$	&$(0.8889, 0.1111)$ \\ 
\hline
$7$	&$(0.8889, 0.1111)$	&$(0.8889, 0.1111)$	&$(0.8889, 0.1111)$ \\ 
\hline
$8$	&$(0.8889, 0.1111)$	&$(0.8889, 0.1111)$	&$(0.8889, 0.1111)$ \\
\hline
\hline
\end{tabular*}
\label{para}
\end{table}

We can also look at this dynamics from the ODEs (ordinary differential equations) by examining the continuous limit.
\begin{align}
    \begin{pmatrix}
\dot{p}_{A} \\ 
\dot{p}_{B}
\end{pmatrix}
=
\begin{pmatrix}
p_{An+1} - p_{An} \\ 
p_{Bn+1} - p_{Bn}
\end{pmatrix}
=
\begin{pmatrix}
-0.1 &0.8 \\
0.1 &-0.8
\end{pmatrix}
\begin{pmatrix}
p_{A} \\ 
p_{B}
\end{pmatrix}.
\end{align}
The ODEs from the two equations are equivalent, because of the constraint $p_A+p_B=1$. If we focus on $p_A$, we get the ODE:
\begin{align}
    \dot{p}_{A}=0.8-0.9p_{A}=f(p_A).
\end{align}
From the plot of $p_A$-$f(p_A)$, it is obvious that $p_A$ will flow towards the stable fixed point $f(p_A^*)=0, p_A^*=8/9$, as the ODE says that $p_A$ should increase if $f(p_A)>0$ below $p_A^*$ and otherwise should decrease if $f(p_A)<0$ above $p_A^*$. Therefore, $p_A$ will always flow towards its fixed point from any initial condition. The final steady state is in line with our linear algebra result.

\begin{figure*}[htb]
\begin{center}
\subfigure[]{\includegraphics[width=0.24\textwidth]{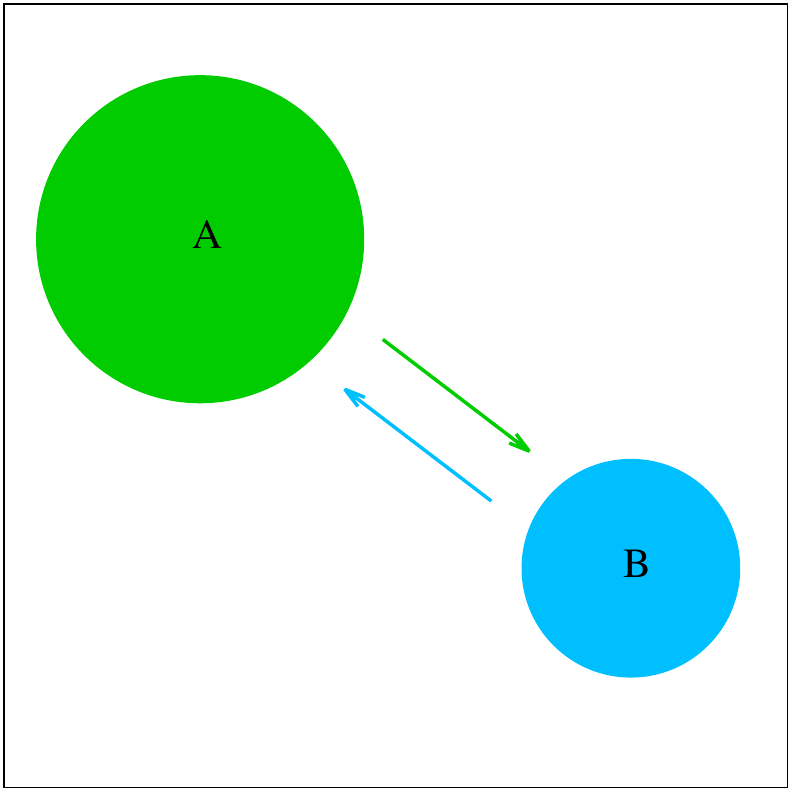} \label{TS}}
\subfigure[]{\includegraphics[width=0.24\textwidth]{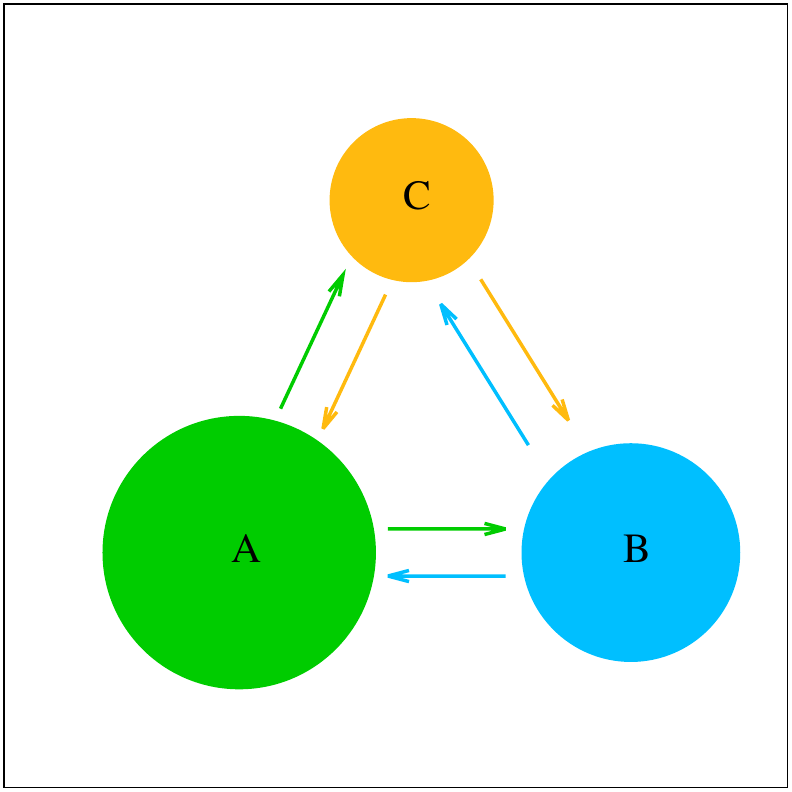} \label{ThreeS1}}
\subfigure[]{\includegraphics[width=0.24\textwidth]{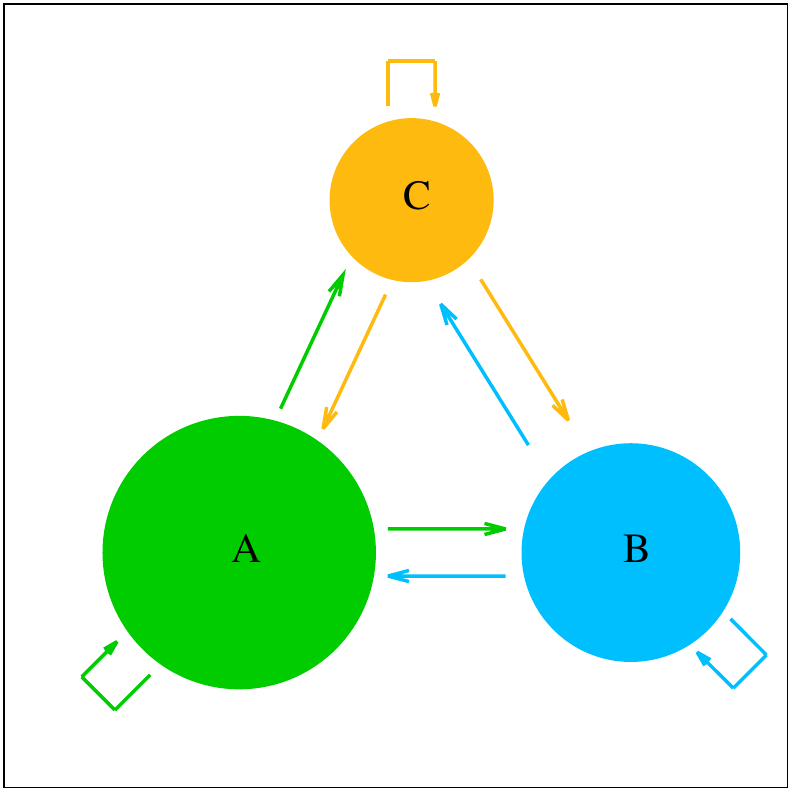} \label{ThreeS2}}
\subfigure[]{\includegraphics[width=0.24\textwidth]{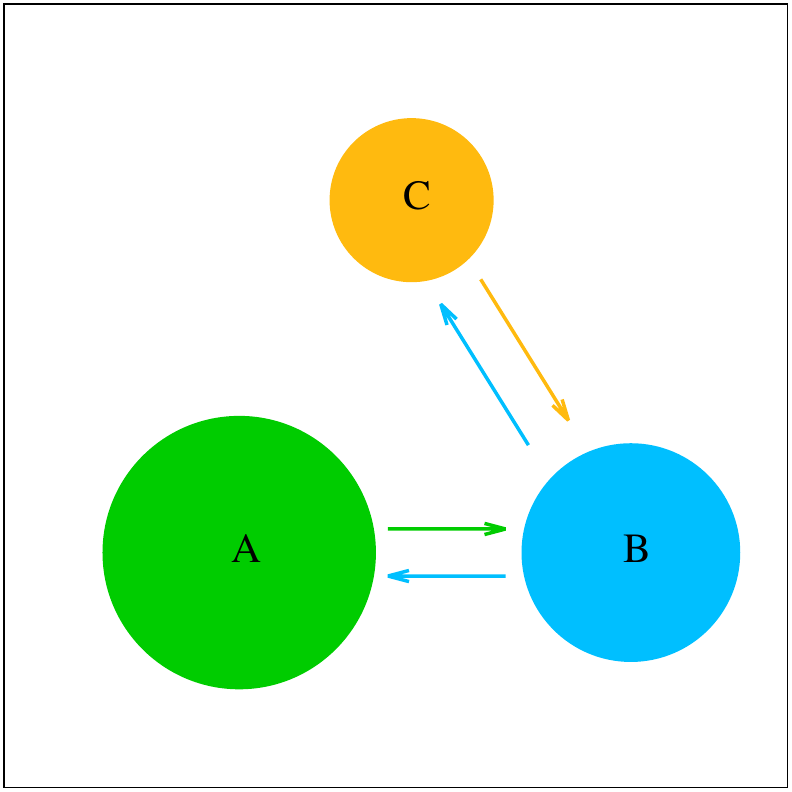} \label{ThreeS3}}
\subfigure[]{\includegraphics[width=0.24\textwidth]{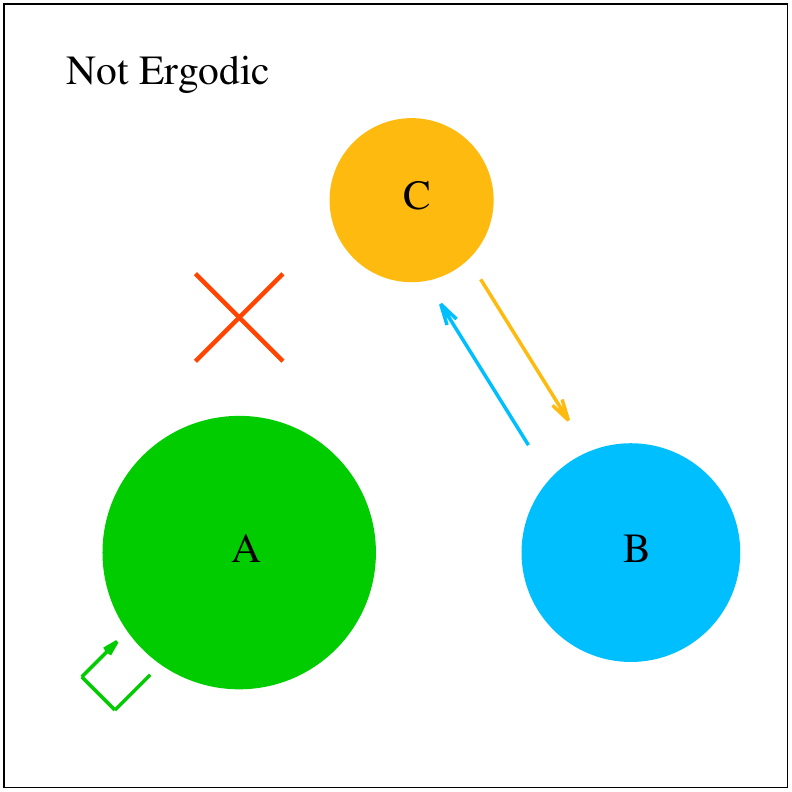} \label{ThreeS4}}
\subfigure[]{\includegraphics[width=0.24\textwidth]{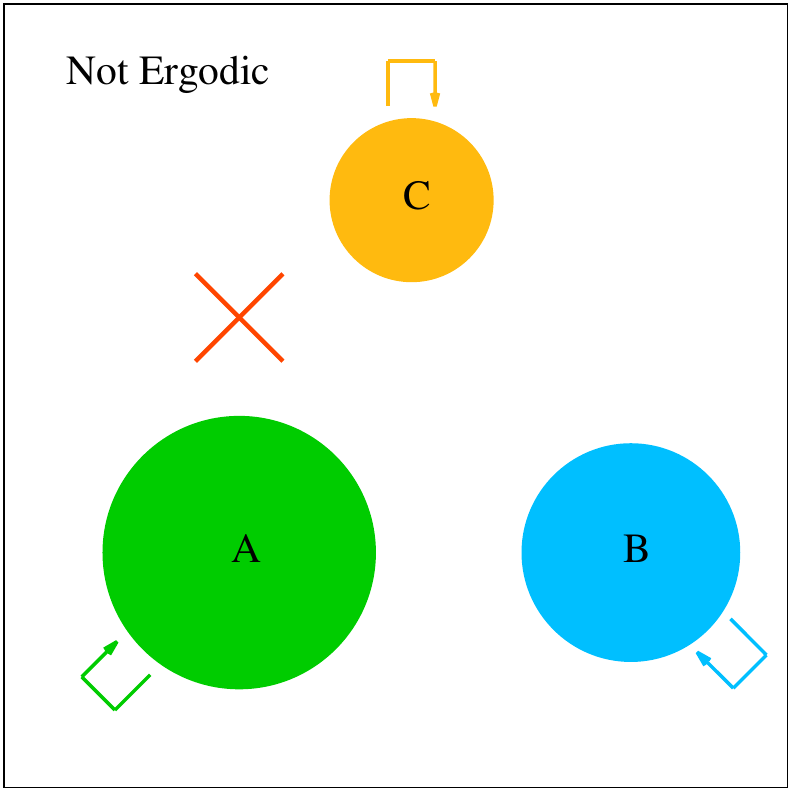} \label{ThreeS5}}
\subfigure[]{\includegraphics[width=0.24\textwidth]{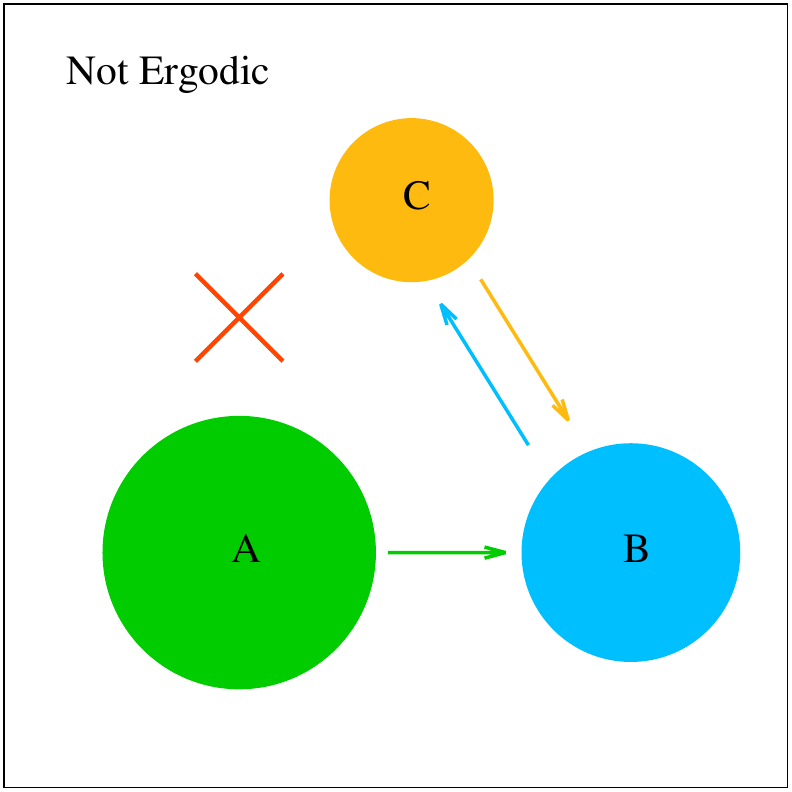} \label{ThreeS6}}
\subfigure[]{\includegraphics[width=0.24\textwidth]{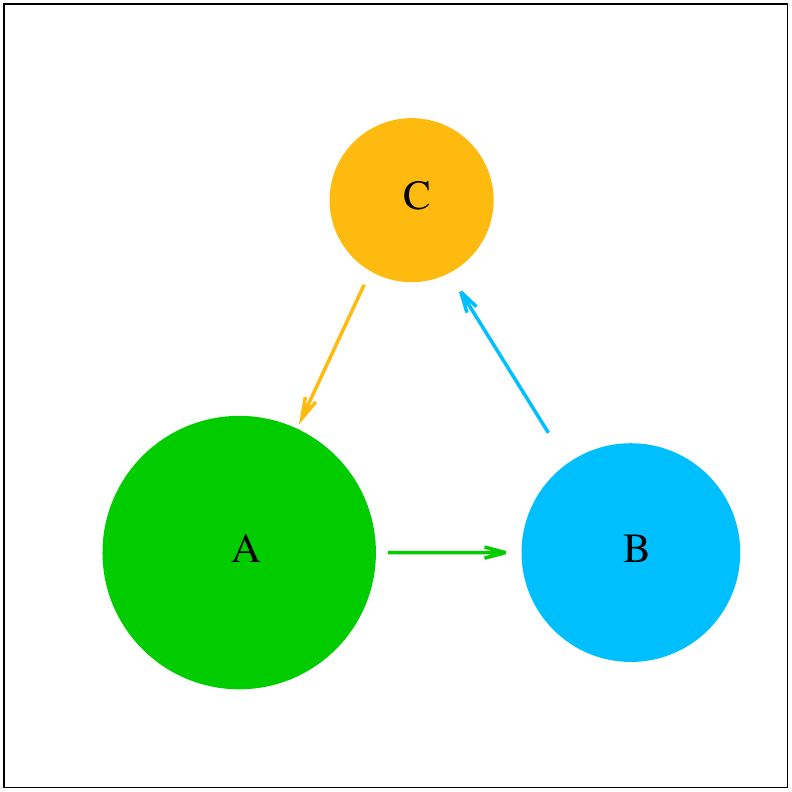} \label{ThreeS7}}
\caption{
Random walk dynamics leads to an equilibrium distribution, providing a novel way of sampling the two-state distribution with suitable random walk rules (a). The same holds for three states (b). In this figure, the arrows represent selection or proposal probabilities, here, we can assume they are uniformly distributed for simplicity, the acceptance probability is then calculated from the detailed balance condition. See the text for details. The proposal scheme is, however, not uniqure. In (c) a state can select itself and in (d) A and C are not directly connected, all of these schemes are valid. However, the selection schemes in (e-g) are not ergodic, e.g., if we start a state at C, there is no chance for A to appear. It is possible to satisfy global balance, breaking the detailed balance, by forming loops of states, as illustrated in (h).
}
\label{TS2}
\end{center}
\end{figure*}

It is important to realize that this dynamics provides \textbf{a novel way of sampling} a distribution by doing \textbf{random walks}. If we intialize a state with $0$ for city A or $1$ for city B, then we can keep updating the state following the transition rules or the random walk rules, then we expect $0$ to appear with probability $8/9$ and $1$ to appear with probability $1/9$. This approach is clearly different from the inversion method based on the CDF (cumulative distribution function) of the distribution. We designed this population dynamics as our introductory example because it is both simple and very close to the nature of the MC random walk.

The particular fixed point $(p_A^*, p_B^*)=(8/9, 1/9)$ is obviously a consequence of the specific transition rules. Now we try to generalize the method and reverse the problem: Given a desired distribution $(p_A^*, p_B^*)$, how to design the transition rules such that the flow will converge to the desired distribution?

We just repeat the calculation, but now we keep generally that $p_{A\rightarrow B}=p$ and $p_{B\rightarrow A}=q$. The dynamics reads:
\begin{align}
    \begin{pmatrix}
p_{An+1} \\ 
p_{Bn+1}
\end{pmatrix}
=
\begin{pmatrix}
1-p &q \\
p &1-q
\end{pmatrix}
\begin{pmatrix}
p_{An} \\ 
p_{Bn}
\end{pmatrix}.
\end{align}
Using the simpler ODE analysis, we have:
\begin{align}
    \begin{pmatrix}
\dot{p}_{A} \\ 
\dot{p}_{B}
\end{pmatrix}
&=
\begin{pmatrix}
-p &q \\
p &-q
\end{pmatrix}
\begin{pmatrix}
p_{A} \\ 
p_{B}
\end{pmatrix}, \\
p_A^* p &= p_B^* q, \\
p_A^* p_{A\rightarrow B} &= p_B^* p_{B\rightarrow A}.
\label{detailbalance2}
\end{align}
The result is quite interesting, it has a very intuitive interpretation. Equation~\eqref{detailbalance2} says to achieve the desired equilibrium distribution, we should \textbf{balance} the probability flows: the flow from A to B should be the same as the flow from B to A. Precisely, the weight of A times the transition probability from A to B should equal the weight of B times the transition probability from B to A. This equation provides the basis to design the transition probabilities.

The choice of the transition probabilities, however, is not unique. Considering that $\frac{p_{A\rightarrow B}}{p_{B\rightarrow A}}
=\frac{p_B^*}{p_A^*}$, it is one convention to set:
\begin{align}
    p_{A\rightarrow B}
=\frac{p_B^*}{p_A^*+p_B^*}=p_B^*, \
p_{B\rightarrow A}
=\frac{p_A^*}{p_A^*+p_B^*}=p_A^*.
\label{heatbath}
\end{align}
This choice is known as the \textbf{heat bath algorithm}, it is also used in the Glauber dynamics \cite{Glauber:MC}.
Another idea is to maximize the flowing dynamics, i.e., we try to set the largest transition probability to $1$. In the case $p_A^*>p_B^*$, we rewrite the balance condition as $p_{A\rightarrow B} = \frac{p_B^*}{p_A^*} p_{B\rightarrow A}$. Then, we can set $p_{A\rightarrow B}
=\frac{p_B^*}{p_A^*}, \quad
p_{B\rightarrow A}
=1$. Otherwise, if $p_A^*<p_B^*$, we rewrite the balance condition as $\frac{p_A^*}{p_B^*}p_{A\rightarrow B} =  p_{B\rightarrow A}$. Then, we set $p_{A\rightarrow B}
=1, \quad
p_{B\rightarrow A}
=\frac{p_A^*}{p_B^*}$. From the perspective of A, we certainly move to B if B has a larger weight or otherwise move to B with a probability $p_B^*/p_A^*$. Similarly, from the perspective of B, we move to A for sure if A has a larger weight or otherwise we move to A with a probability $p_A^*/p_B^*$. Because of this similarity, the transition probability from an initial state $i$ to a final state $f$ can be written compactly as:
\begin{align}
p_{i\rightarrow f} = \min(p_f/p_i, 1).
\label{Met}
\end{align}
This choice of setting the transition probabilities is the well-known \textbf{Metropolis algorithm}. Now, we can sample the Bernoulli distribution using the Markov Chain Monte Carlo method. Here, the chain means that we are generating \textbf{a chain of states} from the random walk dynamics. In this work, we focus on the Metropolis algorithm, which is one of the most popular algorithms in Monte Carlo simulations \cite{metropolis:53}.

\begin{algorithm}[htb]
\caption{A MCMC algorithm for the two-state system}            
\label{algmcmc}                       
\begin{algorithmic}
\REQUIRE System $x$, Monte Carlo steps $N$, weights of the two states $w_0, w_1$. 
\ENSURE MCMC states.
\STATE Initialize state $x=0$. //You can also start from $x=1$ or randomly from $0$ or $1$.
\FOR{$i=1:N$}       
\IF{$x==0$}                        
\STATE Reset $x=1$ if $rand()<w_1/w_0$.          
\ELSIF{$x==1$}
\STATE Reset $x=0$ if $rand()<w_0/w_1$.
\ENDIF
\STATE Make measurement of states, e.g., you can sum $x_i$ for the sample average.
\ENDFOR
\end{algorithmic}
\end{algorithm}

Finally, we summarize the Metropolis algorithm for the two-state system in Alg.~\ref{algmcmc}. For each cycle we try to update a state, we can it a \textbf{Monte Carlo step}. The update itself is also known as the \textbf{Monte Carlo move}, the repeated application of which generates a chain of states as a function of the Monte Carlo step or time. It is worth mentioning that if a transition is rejected, the present state is counted again, otherwise, we would simply register states like $010101...$ which is clearly not correct. It is frequently considered as a tricky concept that a rejected state counts again in the Markov Chain. Interestingly, there is essentially no chance for this confusion to arise in our context, highlighting the importance of the simple two-state system.

\section{Three states}
We have discussed the random walk sampling for the two-state system, it is interesting whether this technique works for more states. Similarly, the first sensible step could be to check whether this dynamics still runs into an equilibrium state for a particular example. To this end, we study the following specific three-state dynamics:
\begin{align}
\begin{pmatrix}
p_{An} \\ 
p_{Bn} \\
p_{Cn}
\end{pmatrix}
=
\begin{pmatrix}
0.2 &0.1 &0.3 \\
0.4 &0.1 &0.2 \\
0.4 &0.8 &0.5
\end{pmatrix}
\begin{pmatrix}
p_{A_{n-1}} \\ 
p_{B_{n-1}} \\ 
p_{C_{n-1}}
\end{pmatrix}.
\end{align}
These numbers are filled quite arbitrarily, expect that each column sums to $1$. This iteration is more complicated to analyze, but we can easily solve it numerically. From the calculations, we observe the following facts:
\begin{enumerate}
    \item The largest eigenvalue is exactly $1$, and all the other eigenvalues have magnitude less than $1$.
    \item The iteration runs into an equilibrium, from any initial condition, and the equilibrium state is the eigenvector of the eigenvalue $1$.
\end{enumerate}
We can also fill the matrix with $U[0,1]$ random numbers and then we properly normalize each column. The above features appear very robust, suggesting that it is promising to sample a three-state system using random walks. We discuss later that such matrices are called \textbf{stochastic matrices} for describing \textbf{stochastic processes}.

We should try to reverse the problem and design random walk rules for a desired distribution. For a generic equilibrium state $\vec{p}^*=(p_A^*, p_B^*, p_C^*)^T$, the stochastic $S$ matrix should satisfy some constraints:
\begin{align}
    \vec{p}^* &=S\vec{p}^*, \quad p_i^* = \sum_j S_{ij} p_j^*, \label{eve} \\
    \sum_i S_{ij} &= \sum_i p_{j \rightarrow i} =1.
\end{align}
The $\vec{p}^*$ is an eigenstate of $S$ with eigenvalue $1$, we should see later that it is the eigenstate of the eigenvalue $1$ and all the other eigenvalues have magnitude strictly less than $1$ if the random walk or $S$ is healthy. The $S$ matrix has $9$ elements, it is not very straightforward to design $S$, it is much easier to design random walk rules directly from the balance condition. We have done so for the two-state system, note that we designed the random walk rules from the balance equation $p_A^* p_{A\rightarrow B} = p_B^* p_{B\rightarrow A}$. From the random walk rules, we can straightforwardly calculate the transition probabilities and therefore the $S$ matrix. For the heat bath algorithm (Eq.~\eqref{heatbath}) and the Metropolis algorithm (Eq.~\eqref{Met}), the respective $S$ matrices are:
\begin{align}
S_{\mathrm{HB}}&=
\begin{pmatrix}
p_A^* &p_A^* \\
p_B^* &p_B^*
\end{pmatrix}, \\
S_{\mathrm{Met}}&=
\begin{pmatrix}
0 &p_A^*/p_B^*\\
1 &1-p_A^*/p_B^*
\end{pmatrix},
\end{align}
where we have assumed the case $p_A^*<p_B^*$ for simplicity. 
It is extremely important to emphasize that we are here writing down the $S$ matrices \textit{after} we designed the random walk rules, which come from the balance of the probability flows. While the $S$ matrix is extremely helpful in analyzing the random walk dynamics, it is not strictly needed when designing a MC random walk or simulation. Indeed, for a large state space, we merely design random walk rules without explicitly writing down the $S$ matrix.

To study the three-state system, one natural generalization is perhaps to extend the balance equation to each pair of states. It might be sensible to expect that if we balance the flows between any pair of states, then the full states will be in equilibrium. This is in fact true, and this condition is known as the \textbf{detailed balance} because the probability flows are all balanced in a detailed pairwise manner. In terms of the $S$ matrix, we require that $S_{ij}p_j^*=S_{ji}p_i^*$. This is a sufficient condition to satisfy the equilibrium condition $\vec{p}^* = S\vec{p}^*$ as:
\begin{align}
p_i^* p_{i\rightarrow j} &= p_j^* p_{j\rightarrow i}, \\
p_i^* S_{ji} &= p_j^* S_{ij}, \\
\sum_j S_{ij} p_j^* &= \sum_j S_{ji} p_i^* = p_i^* \sum_j S_{ji} =p_i^*, \\
\vec{p}^* &=S\vec{p}^*.
\end{align}

Next, we design random walk rules from the detailed balance equations. The simplest scheme perhaps is to design the random walk process in two steps, first a \textbf{proposal step or selection step}, and then an \textbf{acceptance step}. This two-step process significantly simplifies the design of walking rules. The proposal step is quite new here, the motivation for it is that a state herein is typically connected with more than one state. At each state, we first distribute a probability distribution where we propose to move. Note that this step can be designed with a good degrees of freedom. For example, the state A can propose to move to B or C with an equal probability $1/2$, and similarly for B and C upon cyclic permutations, as illustrated in Fig.~\ref{ThreeS1}. An alternative choice is that each state selects A, B, C with an equal probability $1/3$, i.e., a state can select itself, as shown in Fig.~\ref{ThreeS2}. At each state, the proposal probability of this step is normalized to $1$, therefore, this is a well-defined random process. One may wonder why we can somewhat set these numbers almost at will, the reason is that we have not set the \textbf{acceptance probabilities}. For the designed \textbf{proposal probabilities}, we can still tune the acceptance probabilities to satisfy the detailed balance:
\begin{align}
p_{i\rightarrow j}&=p_{\mathrm{prop}, i\rightarrow j} p_{\mathrm{accp}, i\rightarrow j}, \quad i\neq j, \label{tran}\\
p_i^* p_{\mathrm{prop}, i\rightarrow j} p_{\mathrm{accp}, i\rightarrow j}
&=
p_j^* p_{\mathrm{prop}, j\rightarrow i} p_{\mathrm{accp}, j\rightarrow i}.
\label{detailbalance3}
\end{align}
We can maximize the acceptance probabilities using the Metropolis algorithm:
\begin{align}
p_{\mathrm{accp}, i\rightarrow j}
= \min\left[\frac{p_j^* p_{\mathrm{prop}, j\rightarrow i}}{p_i^* p_{\mathrm{prop}, i\rightarrow j}}, \ 1\right].
\label{paccp3}
\end{align}
This equation is no different from Eq.~\eqref{Met} of the two-state system if we view the product of the weight and the proposal probability as a whole. This makes intuitive sense, because $p_i^* p_{\mathrm{prop}, i\rightarrow j}$ is the effective weight that actually tries to flow from $i$ to $j$ and vice versa. Similarly, Eq.~\eqref{detailbalance3} is also not very different from Eq.~\eqref{detailbalance2} in terms of the effective weight.

\begin{algorithm}[htp]
\caption{A MCMC algorithm for the three-state system}  
\label{algmcmc2}                       
\begin{algorithmic}
\REQUIRE System $x$, Monte Carlo steps $N$, weights of the three states $w_0, w_1, w_2$. We choose each neighbour with $p=1/2$. 
\ENSURE MCMC states.
\STATE Initialize state $x=0$. //You can also start from $x=1, 2$ or randomly from the three states.
\FOR{$i=1:N$}       
\IF{$x==0$}
\IF{$rand()<=0.5$}
\STATE Reset $x=1$ if $rand()<w_1/w_0$.  
\ELSE
\STATE Reset $x=2$ if $rand()<w_2/w_0$.
\ENDIF
\ELSIF{$x==1$}
\IF{$rand()<=0.5$}
\STATE Reset $x=0$ if $rand()<w_0/w_1$.  
\ELSE
\STATE Reset $x=2$ if $rand()<w_2/w_1$.
\ENDIF
\ELSIF{$x==2$}
\IF{$rand()<=0.5$}
\STATE Reset $x=0$ if $rand()<w_0/w_2$.  
\ELSE
\STATE Reset $x=1$ if $rand()<w_1/w_2$.
\ENDIF
\ENDIF
\STATE Make measurement of states, e.g., you can sum $x_i$ for the sample average.
\ENDFOR
\end{algorithmic}
\end{algorithm}

We now try a specific example and practise our skills for the three-state system. Suppose that we would like to sample the distribution $\vec{p}^*=(0.6, 0.25, 0.15)^T$, and we start with the selection scheme that a state chooses each of its two neighbours with equal probability $1/2$. The detailed balance conditions and acceptance probabilities are:
\begin{align}
    p_A^* \times \frac{1}{2} \times p_{\mathrm{accp}, A\rightarrow B}
&=
p_B^* \times \frac{1}{2} \times p_{\mathrm{accp}, B\rightarrow A}, \\
p_{\mathrm{accp}, A\rightarrow B}&=\frac{p_B^*}{p_A^*}, \\
p_{\mathrm{accp}, B\rightarrow A}&=1, \\
    p_A^* \times \frac{1}{2} \times p_{\mathrm{accp}, A\rightarrow C}
&=
p_C^* \times \frac{1}{2} \times p_{\mathrm{accp}, C\rightarrow A}, \\
p_{\mathrm{accp}, A\rightarrow C}&=\frac{p_C^*}{p_A^*}, \\
p_{\mathrm{accp}, C\rightarrow A}&=1, \\
    p_B^* \times \frac{1}{2} \times p_{\mathrm{accp}, B\rightarrow C}
&=
p_C^* \times \frac{1}{2} \times p_{\mathrm{accp}, C\rightarrow B}, \\
p_{\mathrm{accp}, B\rightarrow C}&=\frac{p_C^*}{p_B^*}, \\
p_{\mathrm{accp}, C\rightarrow B}&=1.
\end{align}

Note that this is already sufficient to do the MCMC random walk or sampling, the algorithm is shown in Alg.~\ref{algmcmc2}. Nevertheless, let us keep working and figure out the transition probabilities:
\begin{align}
  p_{A \rightarrow B}&=\frac{1}{2}p_{\mathrm{accp}, A\rightarrow B}=\frac{1}{2}\frac{p_B^*}{p_A^*}, \\
  p_{A \rightarrow C}&=\frac{1}{2}p_{\mathrm{accp}, A\rightarrow C}=\frac{1}{2}\frac{p_C^*}{p_A^*}, \\
  p_{A \rightarrow A}&=\frac{1}{2}\frac{p_A^*-p_B^*}{p_A^*}+\frac{1}{2}\frac{p_A^*-p_C^*}{p_A^*}, \\
p_{B \rightarrow A}&=\frac{1}{2}p_{\mathrm{accp}, B\rightarrow A}=\frac{1}{2}, \\
p_{B \rightarrow C}&=\frac{1}{2}p_{\mathrm{accp}, B\rightarrow C}=\frac{1}{2}\frac{p_C^*}{p_B^*}, \\
p_{B \rightarrow B}&=\frac{1}{2}\frac{p_B^*-p_C^*}{p_B^*}, \\
p_{C \rightarrow A}&=\frac{1}{2}p_{\mathrm{accp}, C\rightarrow A}=\frac{1}{2}, \\
p_{C \rightarrow B}&=\frac{1}{2}p_{\mathrm{accp}, C\rightarrow B}=\frac{1}{2}, \\
p_{C \rightarrow C}&=0.
\end{align}
It is worth noting that A can effectively move to A even if we never proposed so, this can happen when A tried to move to either B or C but the move was not accepted. This is similarly true for the state B. For a pair of distinct states with mutual proposal probabilities, the transition probability is simply the product of the proposal probability and the acceptance probability (Eq.~\eqref{tran}) such as $p_{A\rightarrow B}$ and $p_{C\rightarrow A}$.

It is interesting that these transition probabilities do make a proper $S$ matrix, the sum of each column is $1$. This is not surprising, as our random processes are well defined. As mentioned earlier, the first proposal step itself is well defined. The second acceptance step is also well defined. If a move is accepted, the state is changed. If the move is rejected, the state remains the same. Whatever happens, the state will certainly do something with a total probability $1$. The $S$ matrix is summarized here, and one can run the iteration numerically and confirm that it does converge to the desired distribution from any valid initial state.
\begin{align}
\begin{pmatrix}
p_{An} \\ 
p_{Bn} \\
p_{Cn}
\end{pmatrix}
=
\begin{pmatrix}
0.6667    &0.5000    &0.5000 \\
0.2083    &0.2000    &0.5000 \\
0.1250    &0.3000    &0
\end{pmatrix}
\begin{pmatrix}
p_{A_{n-1}} \\ 
p_{B_{n-1}} \\ 
p_{C_{n-1}}
\end{pmatrix}.
\end{align}

As mentioned earlier, the proposal probability can be tuned quite flexibly, and the detailed balance is taken care of by the acceptance probability. Indeed, let us try a second proposal scheme in which we choose each state with equal probability $p=1/3$ including the present state itself. The detailed balance equations and the acceptance probabilities are summarized below:
\begin{align}
    p_A^* \times \frac{1}{3} \times p_{\mathrm{accp}, A\rightarrow B}
&=
p_B^* \times \frac{1}{3} \times p_{\mathrm{accp}, B\rightarrow A}, \\
p_{\mathrm{accp}, A\rightarrow B}&=\frac{p_B^*}{p_A^*}, \\
p_{\mathrm{accp}, B\rightarrow A}&=1, \\
    p_A^* \times \frac{1}{3} \times p_{\mathrm{accp}, A\rightarrow C}
&=
p_C^* \times \frac{1}{3} \times p_{\mathrm{accp}, C\rightarrow A}, \\
p_{\mathrm{accp}, A\rightarrow C}&=\frac{p_C^*}{p_A^*}, \\
p_{\mathrm{accp}, C\rightarrow A}&=1, \\
    p_B^* \times \frac{1}{3} \times p_{\mathrm{accp}, B\rightarrow C}
&=
p_C^* \times \frac{1}{3} \times p_{\mathrm{accp}, C\rightarrow B}, \\
p_{\mathrm{accp}, B\rightarrow C}&=\frac{p_C^*}{p_B^*}, \\
p_{\mathrm{accp}, C\rightarrow B}&=1.
\end{align}
Interestingly, there is hardly any change, as the selection probabilities herein are very symmetric. Particularly, the acceptance probabilities are identical as before. However, these two sets of random walk rules are different, as the transition probabilities are indeed different:
\begin{align}
  p_{A \rightarrow B}&=\frac{1}{3}p_{\mathrm{accp}, A\rightarrow B}=\frac{1}{3}\frac{p_B^*}{p_A^*}, \\
  p_{A \rightarrow C}&=\frac{1}{3}p_{\mathrm{accp}, A\rightarrow C}=\frac{1}{3}\frac{p_C^*}{p_A^*}, \\
  p_{A \rightarrow A}&=\frac{1}{3}+\frac{1}{3}\frac{p_A^*-p_B^*}{p_A^*}+\frac{1}{3}\frac{p_A^*-p_C^*}{p_A^*}, \\
p_{B \rightarrow A}&=\frac{1}{3}p_{\mathrm{accp}, B\rightarrow A}=\frac{1}{3}, \\
p_{B \rightarrow C}&=\frac{1}{3}p_{\mathrm{accp}, B\rightarrow C}=\frac{1}{3}\frac{p_C^*}{p_B^*}, \\
p_{B \rightarrow B}&= \frac{1}{3}+ \frac{1}{3}\frac{p_B^*-p_C^*}{p_B^*}, \\
p_{C \rightarrow A}&=\frac{1}{3}p_{\mathrm{accp}, C\rightarrow A}=\frac{1}{3}, \\
p_{C \rightarrow B}&=\frac{1}{3}p_{\mathrm{accp}, C\rightarrow B}=\frac{1}{3}, \\
p_{C \rightarrow C}&=\frac{1}{3}.
\end{align}
This $S$ matrix is therefore different from the previous one, but the dynamics also converges to the desired distribution:
\begin{align}
\begin{pmatrix}
p_{An} \\ 
p_{Bn} \\
p_{Cn}
\end{pmatrix}
=
\begin{pmatrix}
0.7778    &0.3333    &0.3333 \\
0.1389    &0.4667    &0.3333 \\
0.0833    &0.2000    &0.3333
\end{pmatrix}
\begin{pmatrix}
p_{A_{n-1}} \\ 
p_{B_{n-1}} \\ 
p_{C_{n-1}}
\end{pmatrix}.
\end{align}
Such selection parameters can be typically tuned in wide ranges, and one can tune them to optimize the algorithm. For example, it is also possible for each state to choose itself with $p=0.98$ and randomly choose a neighbour with only a probability $p=0.01$. While there is nothing wrong with this selection rule, it is likely not a very good choice. We shall discuss the error analysis later, for now we focus on how MCMC works.

There is actually another condition for the design to work, the \textbf{ergodicity} condition, in addition to the balance condition. This principle says that the flow should be healthy, one can in principle reach any state from any initial state. For example, if C always selects itself, this does not work. As if we initialize the state at C, it remains there forever. If A and B always propose each other, this is also not valid, as there is no chance to flow to C starting from either A or B. In addition, if A proposes either B or C, but no state proposes to flow to A, this is also not valid as if we do not start from A, we will never reach A. However, the ergodicity does not require all states to be fully pairwise connected. If A and B can flow towards each other, and similarly for B and C, this is fine. Despite A cannot flow to C in a single step, it can nevertheless reach C by first flowing to B and then from B to C. Some typical examples are depicted in Fig.~\ref{TS2}. Therefore, the ergodicity requires that the states are all connected in a connected network, there should be no disconnected or unreachable clusters of states.

Let us investigate the new selection scheme that the states are not all pairwise connected. Precisely, both states A and C select the central state B, and B randomly selects either A or C with equal probability $p=0.5$. The detailed balance equations and the acceptance probabilities read:
\begin{align}
    0.6 \times 1 \times p_{\mathrm{accp}, A\rightarrow B}
&=
0.25 \times 1/2 \times p_{\mathrm{accp}, B\rightarrow A}, \\
p_{\mathrm{accp}, A\rightarrow B} &=0.125/0.6, \\ p_{\mathrm{accp}, B\rightarrow A} &=1, \\
0.125 p_{\mathrm{accp}, B\rightarrow C}
&=
0.15 p_{\mathrm{accp}, C\rightarrow B}, \\
p_{\mathrm{accp}, B\rightarrow C}&=1, \\ 
p_{\mathrm{accp}, C\rightarrow B}&=0.125/0.15.
\end{align}
Note that there is no direct transition between A and C. Interestingly, the move from B to C is always accepted, despite that C has a smaller weight. This is exactly because of the effective weights from the proposal probabilities, C always selects B, but B only selects C half of the times, and therefore B exchanges weights with C with only an effective weight $0.5p_B^*<p_C^*$. Similarly, the proposed move from C to B is only accepted with a probability, despite that B has a larger weight. It is straightforward to figure out the $S$ matrix in exactly the same manner as before, and confirm numerically that the iteration also converges to the desired distribution from any valid initial state. It is highly recommended that one keeps practising the three-state system including the underlying $S$ matrix until one is fully comfortable with the setup.

Finally, it is worth noting that the selection probability and acceptance probability setup also works for the two-state system, despite this system is sufficiently simple such that the design is hardly necessary. Nevertheless, this is also a good practice to see the coherence, and we provide here a few examples. Consider the distribution $p_A^*=0.6, p_B^*=0.4$, we can work with the following random walk schemes: (1) Both A and B propose to move to each other with probability $1$ (2) A selects A or B randomly with $p=0.5$, and so does B. (3) A selects A or B randomly with $p=0.5$, but B always selects A. Here, one can figure out the acceptance probabilities, compute their S matrices, and then confirm that all of these iterations converge properly. In addition, one should check that these S matrices are reasonable, e.g., each column sums to $1$ and the desired distribution is the eigenvector of the eigenvalue $1$ and the other eigenvalue has a magnitude smaller than $1$. Considering that we have already provided a number of examples, we shall not present the details further here for clarity.

\section{Global balance and stochastic matrix}

The detailed balance is a sufficient but not necessary condition for the stationary condition $\vec{p}^* = S\vec{p}^*$. It is instructive to rewrite this condition as:
\begin{align}
p_i^* &= \sum_j S_{ij} p_j^*, \\
\sum_j S_{ji} p_i^* &= \sum_j S_{ij} p_j^*.
\end{align}
Here, we have multiplied $\sum_j S_{ji}=1$ on the left hand side. This equation has a very intuitive interpretation, it says that the sum of the probability flows out of $i$ equals the sum of the probability flows into $i$, flow in = flow out in equilibrium. This condition is called the \textbf{global balance}, the \textbf{detailed balance} satisfies the global balance by restricting $S_{ji}p_i^*=S_{ij}p_j^*$ for any connected states $i$ and $j$. The detailed balance is a special but rather common setup to achieve the global balance. Nevertheless, one can break the detailed balance but still satisfy the global balance. The idea is to form loops in the random walk, a state has a direction to flow and then loops back to itself \cite{NP:MC}. It is typically harder to design such random walk rules, so one has to be very careful in this and make sure the global balance is indeed satisfied.

To illustrate the global balance, breaking the detailed balance, let us revisit the three-state system and work on the following selection scheme: A always proposes to move to B, and similarly B to C, and then C back to A, as shown in Fig.~\ref{ThreeS7}. The global balance condition leads to:
\begin{align}
    p_A^* p_{\mathrm{accp}, A\rightarrow B}
&=
p_C^* p_{\mathrm{accp}, C\rightarrow A}
=
p_B^* p_{\mathrm{accp}, B\rightarrow C}.
\end{align}
This flowing scheme is clearly ergodic, but a complexity arises that we cannot figure out the acceptance probabilities in a simple pairwise manner.
The first equation says the flow at state A is balanced, the probability flow out (A to B) equals the probability flow in (C to A), and similarly for the flows at states B and C. Now one can appreciate why global balance rules are more difficult to design. In detailed balance, we work with only two states at a time, and the acceptance probabilities can be directly calculated. Here, we have to consider a chain of states, e.g., it is not very obvious how to set the acceptance probabilities efficiently by only looking at the first equation. If we focus on all the equations, we may recognize a solution:
\begin{align}
p_{A \rightarrow B} &=p_{\mathrm{accp}, A\rightarrow B}=p_B^*p_C^*, \\
p_{B \rightarrow C} &=p_{\mathrm{accp}, B\rightarrow C}=p_C^* p_A^*, \\
p_{C \rightarrow A} &=p_{\mathrm{accp}, C\rightarrow A}=p_A^*p_B^*.
\end{align}
We can improve the rules by setting the maximum acceptance probability to $1$ as before. Define $c=\max(p_A^*p_B^*, \ p_B^*p_C^*, \ p_C^*p_A^*)$, we arrive at the following improved solution:
\begin{align}
p_{A \rightarrow B} &=p_{\mathrm{accp}, A\rightarrow B}=p_B^*p_C^*/c, \\
p_{B \rightarrow C} &=p_{\mathrm{accp}, B\rightarrow C}=p_C^* p_A^*/c, \\
p_{C \rightarrow A} &=p_{\mathrm{accp}, C\rightarrow A}=p_A^*p_B^*/c.
\end{align}
By analyzing their underlying $S$ matrices, we can confirm that both sets of rules work and the latter one does appear to converge faster than the former one. The detailed balance is broken, and this striking concept of forming loops of states is illustrated concretely in the context of the three-state system.

We summarize the two key elements of the MCMC method:
\begin{enumerate}
    \item \textbf{Ergodicity}. We can in principle move to any state from any initial state, i.e., every state should be accessible in principle, ideally in a few finite steps.
    \item \textbf{Balance condition}. The total flow in probability and the total flow out probability should balance at any state.  We work with the detailed balance most of the time, but only the global balance is required.
\end{enumerate}

It can be shown mathematically that if these two conditions are satisfied, the sampling of the random walk will converge exponentially to the desired distribution, guaranteed. This follows from the properties of stochastic matrix. The stochastic process and the stochastic matrix themselves are important fields of mathematics \cite{Meyer:LA,Liu2004}, we are therefore not going to discuss them in depth, but we can give a very brief overview to get a sense of some of the most important and pertinent properties to appreciate the validity of our Monte Carlo random walk rules. 

A vector $\vec{p}$ is a \textbf{probability vector}, describing a distribution, if its elements are nonnegative and sum to $1$. A matrix $S$ is a \textbf{stochastic matrix} if each column is a probability vector, i.e., the sum of each column is $1$ (conservation of probability). Note that $S_{ij}$ represents the transition probability from the state $j$ to $i$. Interestingly, the probability and transition probability are respectively replaced by the probability amplitude and transition probability amplitude in quantum mechanics, sharing a similar structure.

It is not difficult to prove the following properties of stochastic matrices:
\begin{enumerate}
\item If $S$ is a stochastic matrix and $\vec{p}$ is a probability vector, then $\vec{q}=S\vec{p}$ is another probability vector.
\item If both $S$ and $T$ are stochastic matrices, so is $ST$. Then $S^n$ is also a stochastic matrix. Expand $\vec{p}$ in the eigenvectors of $S$, we argue that the eigenvalue magnitudes of $S$ are bounded within $1$, i.e., $|\lambda_i|\leq 1$. Otherwise, the probability vector would grow without bound upon repeated operation of $S$. In addition, $|\lambda|=1$ should exist, otherwise, the probability vector would decay to $0$ under evolution.
\item It is not hard to prove that $det(S-I)=0$, therefore, the eigenvalue $\lambda=1$ exists for any stochastic matrix.
\item Multiply $S \vec{v}=\lambda \vec{v}$ by $(1, 1, ..., 1)$, we see that all the $\lambda \neq 1$ eigenstates are unphysical.
\end{enumerate}

The first property can be proved straightforwardly by definition. First, it is obvious that each element of $\vec{q}$ is nonnegative. In addition, $\sum_i q_i = \sum_i \sum_j S_{ij} p_j = \sum_j p_j \sum_i S_{ij}=1$. It follows that $ST$ is a stochastic matrix, as $S$ times each column of $T$ is also a probability vector. Because the probability is conserved, the largest eigenvalue in magnitude should be no larger than $1$ to prevent growth. Similarly, it should be no less than $1$ to prevent decay. The largest magnitude of the eigenvalues is therefore $1$. The eigenvalue $\lambda=1$ is always present, and indeed we have seen it repeatedly in our numerous two-state and three-state examples. To prove this, we sum all of the rows to the first row, then the first row is identically $0$ because each column of $S$ sums to $1$ which cancels the diagonal $-1$ from the $-I$ matrix. For the last property, we get $(\lambda-1)\sum_i v_i=0$. If $\lambda=1$, there is no restriction on $\vec{v}$. If, however, $\lambda \neq 1$, we must have $\sum_i v_i=0$. Such an eigenvector cannot be physical as a probability vector cannot sum its elements to $0$.

It should be noted that not all stochastic matrices are relevant to the Monte Carlo method, the eigenvalue $\lambda=1$ is not necessarily unique and the eigenvalue $\lambda=-1$ can also exist. However, such stochastic matrices are quite pathological. For example, the identity matrix is stochastic, it does not do anything. Such ``random wak'' dynamics is clearly not egordic. On the other hand, the matrix $S=[0, 1; 1, 0]$ has eigenvalues $\lambda=\pm 1$, the component of the eigenvector of $-1$ in an initial state will never decay to $0$ upon iteration. However, if we check this matrix carefully, we find that this matrix merely swaps the probability elements back and forth, which is quite boring. Therefore, this dynamics is not a random walk, it is in fact fully deterministic. If we initialize the system at state A at $t=0$, we can never find the state B at any even time steps. It is not particularly wrong though, it is in a sense egoridic (it can reach both A and B) and satisfies the detailed balance of $p_A^*=0.5, p_B^*=0.5$, the eigenvector of the eigenvalue $1$. The states $ABABAB...$ is a correct ``sampling'' of this distribution in a pathological way. Fortunately, we are not particularly interested in these types of ``random walks'', and we should ignore such rather special cases in the future. To this end, we introduce the regular stochastic matrix for a genuinely healthy random walk. 

If $R$ or $R^n$ (for a finite $n$) is a full matrix with no $0$ elements, $R$ is a \textbf{regular stochastic matrix}. Then the walk is a genuinely healthy random walk, and it is able to access any state from any other state in finite steps. We state without proving the following theorem: \textbf{A regular stochastic matrix has one and only one eigenvalue $\lambda=1$, and has all other eigenvalues $|\lambda_i|<1$. The probability vector converges exponentially to the eigenvector of the eigenvalue $\lambda=1$, which is the equilibrium distribution, from any valid initial state.} See, e.g., \cite{Meyer:LA,Liu2004} for details.

In previous discussions, we see that the desired distribution is an eigenvector of eigenvalue $1$ of the stochastic matrix $S$, see Eq.~\eqref{eve}. Then, the ergodicity ensures that after a few steps or iterations, $S^n$ is a full matrix, and therefore $S$ is a regular stochastic matrix. Therefore, it is guaranteed that the sampling following our MCMC rules will converge exponentially to the desired distribution. After becoming familiar with this connection, one can focus on designing MCMC random walk rules and forget about the $S$ matrix. In the next section, we generalize the setup to many states.



\section{Many states: discrete systems}
Now we can sample two states and three states and know the principles of the MCMC random walk, we are ready to generalize the setup and sample as many states as we wish. The method proceeds in a very similar way, we design random walk rules, i.e., how to move and how to distribute the proposal probabilities, then we figure out the acceptance probabilities from the detailed balance. In this section, we further practise our skills and study the following geometric distribution with an infinity number of states:
\begin{align}
    p_n=q^n(1-q), \ n=0,\ 1,\ 2,\ ...,
    \label{geodist}
\end{align}
where the parameter $0<q<1$. To design the random walk rules, it is helpful to consider the following questions:
\begin{itemize}
\item What is the state space? Answer: $int \ n=0, 1, 2, ...$
\item How to initialize a state? We can initialize a state anywhere we wish, but $n=0$ appears to be a good and natural choice.
\item How to update/move a state? This is the most interesting part, we are basically considering how to modify a state, such that the state may change to another state, that is exactly what we mean by the MC update/move/random walk. Here, we can propose to update $n$ to $n\pm 1$ if possible, note that the $n=0$ has no left neighbour, so the $n=0$ and $n=1$ transitions need a special attention.
\item Is the set of rules ergodic? Please make sure the rules are valid, otherwise, the entire sampling will not work. Our update of moving one step at a time is ergodic.
\item Finally, we fill in the details, i.e., the detailed proposal probabilities, and then calculate the corresponding acceptance probabilities from the detailed balance.
\end{itemize}
It should not be very difficult to design random walk rules, at least elementary ones after going through a few examples, as the fundamental goal here is how to change or modify a state. In our example, it seems simplest to propose a state to randomly move to its nearest neighbour(s). Precisely, if the current state is $0$, we propose it to move to $1$. Otherwise, we propose it to jump to either $n-1$ or $n+1$ each with $p=0.5$. This is ergodic, as we can go from any initial state to any final state by shifting the state around. Here, the design is again not unique as in our two-state or three-state systems. For example, the state $n=0$ may only choose $n=1$ with probability $0.5$ and otherwise selects itself. One can also choose asymmetric proposal probabilities, e.g., a state $n>0$ chooses $n-1$ with a slightly larger probability $0.6$ and chooses $n+1$ with a slightly smaller probability $0.4$. It is also possible to include $\Delta n=\pm 2$ moves again if relevant, and so on.

Next, we calculate the acceptance probabilities from the detailed balance. Here, we proceed with our simple setup and figure out the detailed balance between the special $n=0$ and $n=1$ transition and then a generic transition between $n>0$ and $n+1$. There are actually only two equations to solve, the detailed balance equations and acceptance probabilities read:
\begin{align}
    &p_0 \times  1 \times p_{accp, 0\rightarrow 1} =p_1 \times (1/2) \times p_{accp, 1\rightarrow 0}, \\
    &p_{accp, 0\rightarrow 1}=q/2, \ p_{accp, 1\rightarrow 0}=1, \\
    &p_n \times  (1/2) \times p_{accp, n\rightarrow n+1}=p_{n+1} \times (1/2) \times p_{accp, n+1\rightarrow n}, \\
    &p_{accp, n\rightarrow n+1}=q, \ p_{accp, n+1\rightarrow n}=1.
\end{align}
The transition rules say that if we are proposing to jump from $n$ to $n-1$, the move is always accepted. Otherwise, it is accepted with a probability. The state $n$ tries to diffuse to larger $n$ by the finite acceptance probabilities but is also meanwhile constantly pushed towards $n=0$ by the unit acceptance probability. This makes sense, as the PMF (probability mass function) is a decreasing function of increasing $n$. The walker interestingly stays most of its time in the relatively small $n$ region or the important region, this is known as the \textbf{importance sampling}; cf. the uniform sampling. The analysis of this random walk is not so complicated, the algebra here is in a sense even simpler than that of the three-state system. Keep running, and $n$ will remarkably follow the geometric distribution. The algorithm is summarized in Alg.~\ref{algmcmcgeo}.

\begin{algorithm}[htp]
\caption{A MCMC algorithm for the geometric distribution}
\label{algmcmcgeo}                       
\begin{algorithmic}
\REQUIRE System $n$, Monte Carlo steps $N$, the parameter $q$.
\ENSURE MCMC states.
\STATE Initialize state $n=0$. //You can also start from $n=1, 2$ and so on;
\FOR{$i=1:N$}       
\IF{$n==0$}
\STATE Reset $n=1$ if $rand()<q/2$.  
\ELSE
\IF{$rand()<=0.5$}
\STATE Reset $n=n-1$.  
\ELSE
\STATE Reset $n=n+1$ if $rand()<q$.
\ENDIF
\ENDIF
\STATE Make measurement of states, e.g., you can sum $n_i, n_i^2$ for their sample averages.
\ENDFOR
\end{algorithmic}
\end{algorithm}

Similarly, we can also sample the Poisson distribution $p_n = \frac{\lambda^n\exp(-\lambda)}{n!}, \ n=0, 1, 2, ...$. Note that the two distributions have exactly \textbf{the same state space}. This means that the random walk rules we figured out above can be straightforwardly applied here, except that the transition probabilities should be properly modified as the distribution has been changed. The random walk rules herein are only slightly more complicated as the acceptance probabilities depend on the specific parameter $\lambda$ and also $n$. The details are summarized here:
\begin{align}
    &p_0 \times  1 \times p_{accp, 0\rightarrow 1}=p_1 \times (1/2) \times p_{accp, 1\rightarrow 0}, \\
    &p_{accp, 0\rightarrow 1}=\min[\lambda/2, 1], \\
    &p_{accp, 1\rightarrow 0}=\min[2/\lambda, 1], \\
    &p_n \times  (1/2) \times p_{accp, n\rightarrow n+1}=p_{n+1} \times (1/2) \times p_{accp, n+1\rightarrow n}, \\
    &p_{accp, n\rightarrow n+1}=\min[\lambda/(n+1),1], \\ 
    &p_{accp, n+1\rightarrow n}=\min[(n+1)/\lambda,1].
\end{align}
We can leave the compact expressions as they are. An additional feature is that we do not have to compute the exponential and factorial factors, this is a very generic feature of the MCMC method, \textbf{only relative weight matters}, see Eq.~\eqref{paccp3}. This is very crucial for more complex distributions, e.g., in statistical physics, the full partition function is frequently unknown to compute the absolute probability of a microstate, but the relative weight between two microstates is readily available for the MCMC random walk.

\section{Many states: continuous systems}

To sample a continuous distribution $f(x)$ using random walks, we can view it as the continuous limit of a dense discrete distribution. More precisely, we can view the \textbf{continuous distribution} $f(x)$ as a \textbf{dense discrete one}, $\{x_i, p_i=f(x_i)dx\}$ with a differential step size $dx$, i.e., $x_i$ represents a small interval $[x_i-dx/2, x_i+dx/2]$. This is very much like the discretization procedure of the finite element method. The final result should not depend on the size of $dx$, which should cancel in the calculation as we shall see below.

Consider the standard Gaussian distribution for simplicity, our state is $double \ x$, and we can start from any reasonable number, e.g., $x=0$.
One possible update is to shift our state randomly in the $x$ direction following, e.g., the uniform distribution $U[-h, h]$. Here, $h$ is on the order of the length scale $\sigma=1$. This is a parameter of the random walk rules, and one has to tune the parameter and make sure it is reasonably efficient. It should not be too large, otherwise, we are changing the state too much in a single step. It is likely that we are trying to move it to the tail of the distribution and the acceptance probabilities would be very small. On the other hand, it should not be too small either. If $h$ is very small, the acceptance probabilities are very high, but the state will not move very much after many steps. We find that $h=0.5$ appears to be a good choice.
The detailed balance and the acceptance probability read:
\begin{align}
    f(x_1) dx \frac{dx}{2h} p_{accp, x_1\rightarrow x_2}&=
f(x_2) dx \frac{dx}{2h} p_{accp, x_2\rightarrow x_1}, \\
p_{accp, x_i\rightarrow x_f} &= \min[f(x_f)/f(x_i), 1].
\end{align} 
The $f(x_1)dx$ is the weight of the state $x_1$ and $dx/(2h)$ is the probability to select $x_2$ if $x_2$ is within a distance of $h$ to $x_1$. This is because $x_2$ has a width $dx$ and the total proposal length is $2h$, then the chance to choose $x_2$ is the ratio of the two lengths $dx/(2h)$. Then, we multiply the acceptance probability from $x_1$ to $x_2$. The probability flow in the opposite direction has a similar form, but note that from the perspective of $x_2$, we are sitting at $x_2$ and the chance to select $x_1$ is similarly $dx/(2h)$. The chance $x_1$ selects $x_2$ is therefore the same as the chance $x_2$ selects $x_1$. Interestingly, all the $dx$ factors cancel, and the acceptance probability has no dependence on the differential $dx$. This method is a very generic approach for analyzing continuous systems. The acceptance probability simply says that if the weight of the chosen state is larger, move to it, otherwise, move to it with a probability $f(x_f)/f(x_i)$. Note that we have nowhere specified our Gaussian function form, indeed, the setup is fully generic and it works for any sufficiently well-behaved continuous distribution.

We provide a simple example in statistical mechanics, and consider a single particle in a potential trap $V(x)$ at the inverse temperature $\beta$. It is well-known in statistical mechanics that we only need to sample the potential energy part as the kinetic energy part is independent and exactly solvable. 
The particle appears at $x$ with the Boltzmann weight $\exp(-\beta V(x))$, where $V(x)$ is the potential energy. We have ignored the partition function here, as only relative weight matters in MCMC.
Therefore, we wish to sample according to the weight function $f(x)=\exp(-\beta V(x))$. We already know how to do this from the previous example, we essentially only have to change the Gaussian function to this new function. The acceptance probability is 
$p_{accp, x_i\rightarrow x_f} = \min[f(x_f)/f(x_i), 1]$ or $p_{accp, x_i\rightarrow x_f} = \min[\exp(-\beta (V(x_f)-V(x_i))), 1]$. If the potential energy decreases, we accept the move. Otherwise, we accept it with a probability. Interestingly, this expression frequently appears in statistical mechanics.

It is important to become familiar with designing MCMC random walk rules, and then be able to figure out the acceptance probabilities. This is the central part of the MCMC method. In the next, we come to the data analysis of the correlated data generated by the Markov chain.

\section{Correlated data}
If we plot the sampled states of the MCMC data, e.g., for the geometric distribution, we can immediately notice that they are highly correlated. Correlation means that the data are not independent, compare this with the \textit{iid} data. For example, we only allowed the state to change by $1$ at most in a single step. If we know the present state is $n>0$, we know for sure the next state is either $n-1, n$ or $n+1$. Similarly, a Gaussian number cannot jump by more than the step size $h$ in a single step. 

\begin{figure}[htb]
\begin{center}
\includegraphics[width=\columnwidth]{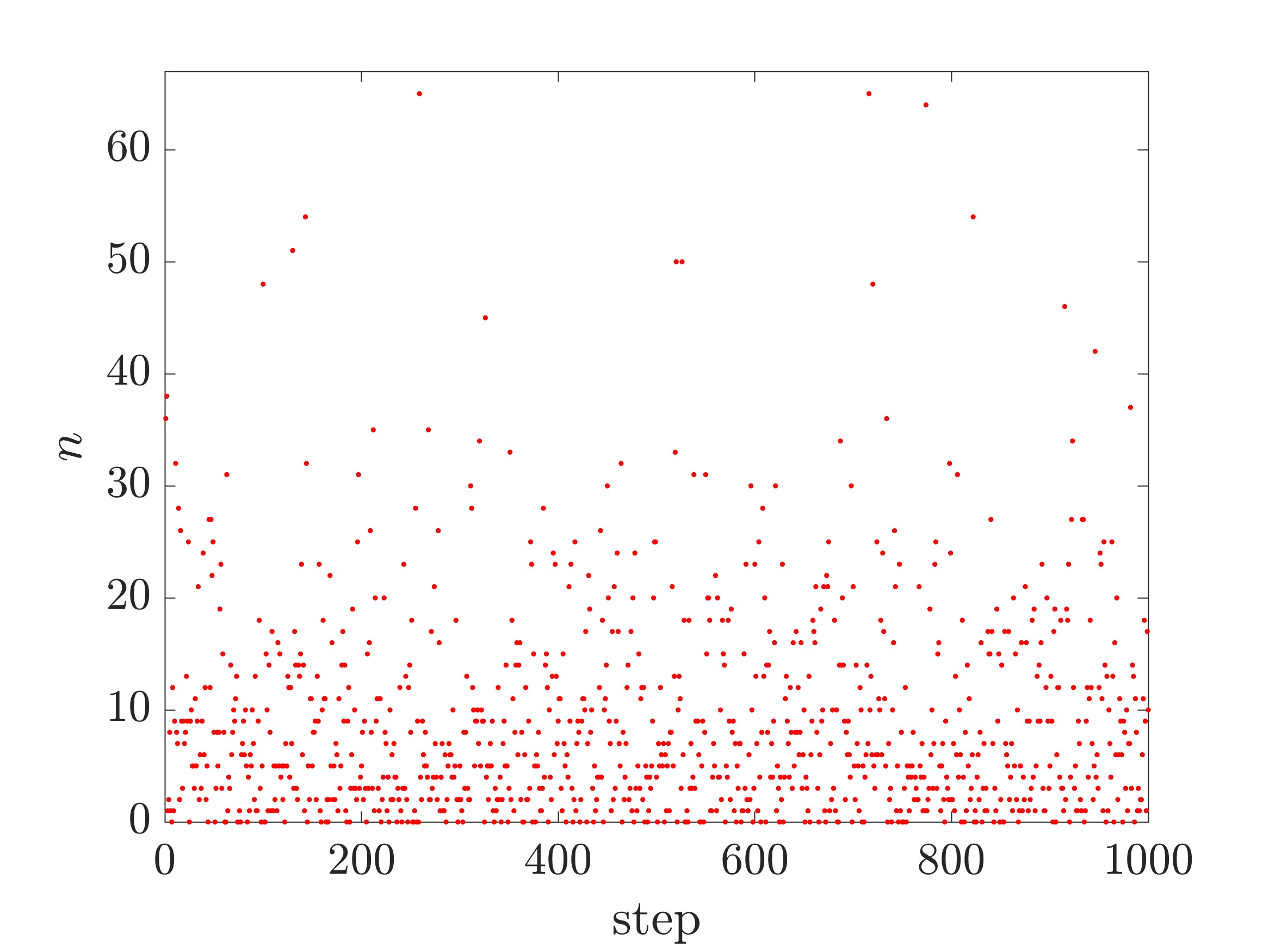}
\includegraphics[width=\columnwidth]{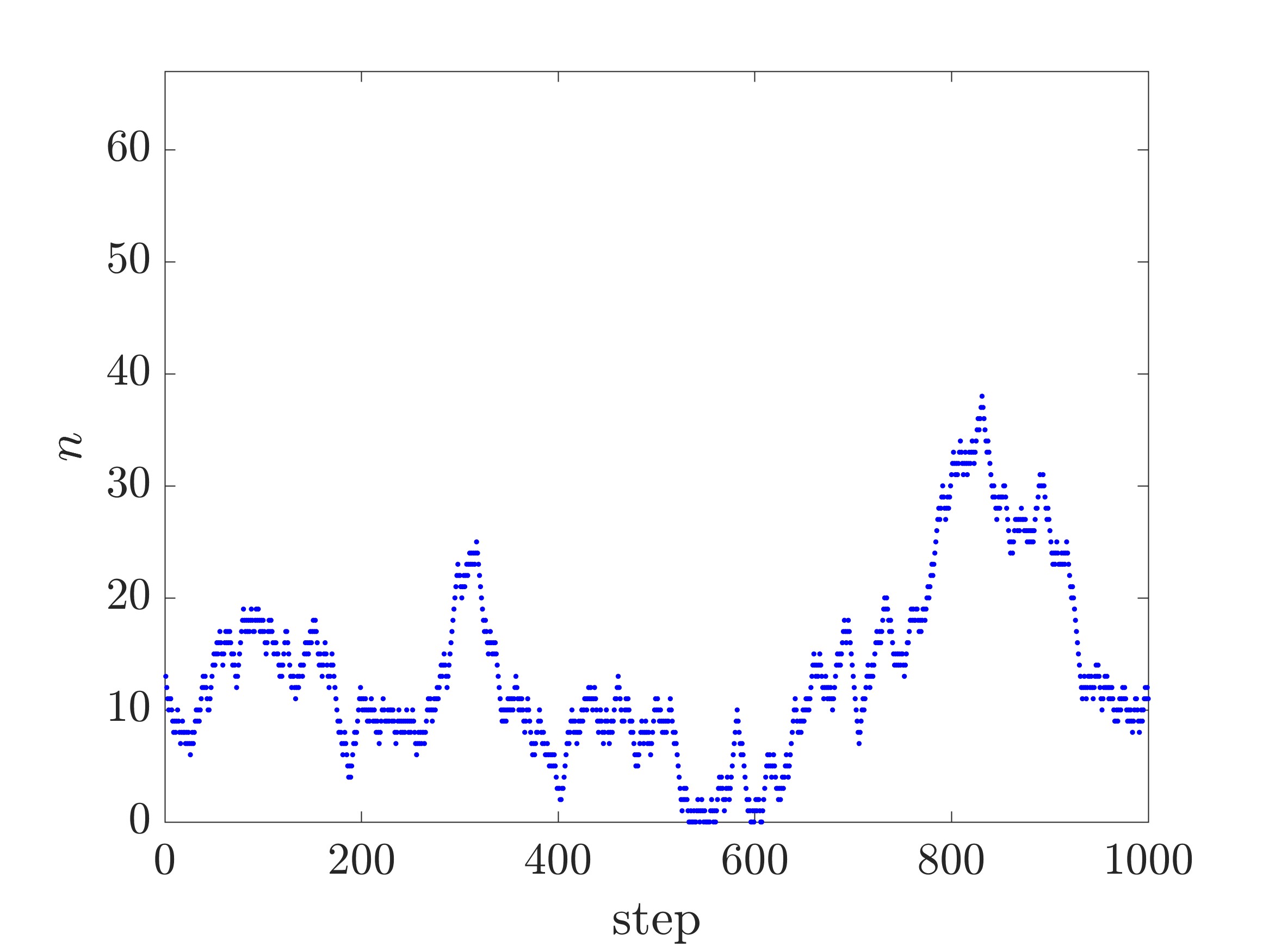}
\caption{
Typical samples of the geometric distribution of Eq.~\eqref{geodist} for $q=0.9$ as a function of the time step. The top panel illustrates uncorrelated \textit{iid} data from the inversion method, and the bottom panel shows the correlated MCMC data. The former bears a random feature, while the latter looks like a trajectory.
}
\label{Corr}
\end{center}
\end{figure}

It is straightforward to visualize the \textbf{correlated data}, and also compare with the \textbf{\textit{iid} data}. It is possible to generate \textit{iid} random numbers of both the geometric distribution and the Gaussian distribution. Typical samples of a geometric distribution are depicted in Fig.~\ref{Corr}. The \textit{iid} data show a very random pattern. Indeed, if we know the value at step $k$, we have no idea which number is coming in the next. By contrast, the MCMC correlated data look like trajectories. These two data types are clearly different. Because correlated data are not \textit{iid} random variables, we cannot directly apply the central limit theorem. Intuitively, we expect correlated data to contain less information as \textit{iid} data for the same data size because it may take some steps to decorrelate the data to get an effectively new sample, now we wish to quantify the correlation and study the data analysis. It is highly important to report a proper errorbar for any estimate in Monte Carlo simulations.

To characterize the data correlation, it is helpful to define the following \textbf{autocorrelation function}:
\begin{align}
C_A(t) &= \frac{\langle A(\tau)A(\tau+t)\rangle_\tau-\langle A(\tau)\rangle_\tau^2}{\langle A^2(\tau)\rangle_\tau-\langle A(\tau)\rangle_\tau^2},
\label{corrfunction}
\end{align}
where $\{A(\tau)\}$ is a list of the recorded data. It is important to mention that \textbf{the time unit here is the period of data collection}. If one records $n$ every step in sampling the geometric distribution, the time unit is one step. If one records the energy of the Ising model every sweep, the time unit is a sweep. If a quantity is measured every $10$ sweeps, then the time unit is $10$ sweeps \cite{Wang:CD}. The first average in Eq.~\eqref{corrfunction} says that we pick a chain of data (from the full data, e.g., a subset of the first $10^6-t$ data if we have a total of $10^6$ data) and also another chain but shifted by a time of $t$, we average their dot product. The second average is just the average of $A$, which is then squared. The denominator is for normalization, it is the variance of $A$ in the chain with no time shift, note that it does not depend on $t$.

The correlation function characterizes how fast the observable $A$ decorrelates in time. We find that $C_A(t=0)=1$ by definition, this physically says that a chain of data is fully correlated with itself, which is clearly correct. In the opposite limit, we expect that $C_A(t=\infty)=0$ as two chains of data that are far separated in time should not correlate in any significant way. In practice, the correlation function is approximately a decaying exponential function. We define two useful times scales from a correlation function:
\begin{align}
\tau_{\rm{int}}^A&=\int_0^{\infty} C_A(t) dt, \\
    \tau_{\rm{exp}}^A:\ C_A(t) &= \exp(-t/\tau_{\rm{exp}}^A).
\end{align}
They are called the \textbf{integrated correlation time} and the \textbf{exponential correlation time}, respectively.
If the decay is perfectly exponential, then $\tau_{\rm{int}}^A=\tau_{\rm{exp}}^A$. In practice, the function might not be a perfect exponential, the two time scales can differ. There is typically a noisy tail as a data set is finite, and the tail fluctuates around $0$ when $t$ is sufficiently large. Here, we should choose a reasonable cutoff, and we \textit{should not integrate or fit the noisy tail} of the correlation function. They are, respectively, estimated as:
\begin{align}
    \tau_{\rm{int}}^A &\approx 1/2+\sum_{i=1}^{\mathrm{cutoff}} C_A(t_i), \\
    \ln(|C_A(t)|) &= -t/\tau_{\rm{exp}}^A+b, \ t \leq t_{\mathrm{cutoff}}.
\end{align}
The former is nothing but a numerical integration by the mid-point rule, and the latter is a log-linear fit. If the exponential fit is reasonable, we should find that $b\approx 0$ as $C_A(t=0)=1$.

The integrated correlation time can be used for estimating the errorbar of the sample mean of an observable. Note that the correlation function depends on the observable. The sample mean itself should not be affected by correlation.
The effective number of independent samples should be smaller than the size of the samples $R$. The estimators of the sample mean and its errorbar along with the \textbf{effective number of samples} are summarized here:
\begin{align}
    \tilde{\overline{o}}&=\frac{\sum \tilde{o}_i}{R}, \\
    \tilde{\sigma}_o^2 &= \frac{\sum (\tilde{o}_i-\tilde{\overline{o}})^2}{R-1}, \\
    \tilde{\sigma}_{\tilde{\overline{o}}} &= \tilde{\sigma}_o\sqrt{\frac{1+2\tau_{\mathrm{int}}^o}{R}}=\frac{\tilde{\sigma}_o}{\sqrt{R_{\mathrm{eff}}}}, \\
    R_{\mathrm{eff}}&=\frac{R}{1+2\tau_{\mathrm{int}}^o}.
\end{align}
Here, the tilde sign denotes a particular set of realized data. The first equation says we average over the correlated data for the sample average, this is meaningful as correlation does not introduce bias, this is therefore the same as the \textit{iid} average. The latter three equations say we can apply the central limit theorem, but the effective number of samples is reduced by a factor of $1+2\tau_{\mathrm{int}}^o$, i.e., the standard deviation of the sample mean is the sample standard deviation divided by the square root of the effective number of samples \cite{Binder:MC}. The expressions may appear like daunting formulas, but the proof is in fact not very complicated:
\begin{align}
    &\mathrm{var}(\overline{x}) = \langle \overline{x}^2 \rangle - \langle \overline{x} \rangle^2, \\
    &= \frac{1}{R^2} \sum_{ij} \left( \langle x_i x_j \rangle - \langle x_i \rangle \langle x_j \rangle \right), \\
    &= \frac{1}{R^2} \Big( \sum_{i} \left( \langle x_i^2 \rangle - \langle x_i \rangle^2 \right) + 2\sum_{i<j} \left( \langle x_i x_j \rangle - \langle x_i \rangle \langle x_j \rangle \right) \Big), \\
    &= \frac{\sigma_x^2}{R^2} \Big( R + 2\sum_{i<j} C(j-i) \Big), \\
    &\approx \frac{\sigma_x^2}{R^2} \left( R + 2R \tau_{\mathrm{int}}^x \right)
=\frac{\sigma_x^2}{R} \left( 1 + 2 \tau_{\mathrm{int}}^x \right).
\end{align}
In the third line, we are separating the diagonal terms and the off diagonal terms. In the fourth line, we utilize the fact that in most rows, $C(j-i)$ decays to $0$ before $j=R$. This is the case if $R\gg \tau_{\mathrm{int}}^x$, only except for the last approximately $\lfloor\tau_{\mathrm{int}}^x\rfloor$ rows, but this number is small compared with $R$.
Here, we see that the effective number of independent samples is indeed smaller than the size of the samples $R$. In the \textit{iid} data, there is no correlation, if we take $\tau_{\mathrm{int}}^x=0$, then $R_{\mathrm{eff}}=R$, we are back to the central limit theorem. Interestingly, the errorbar in both cases scales as $1/\sqrt{R}$.

Note that the correlation time depends on the observable $A$, this means to apply the formulas, we have to calculate the correlation functions of all observables of interest. For example, if we want to estimate the average energy, we should compute the correlation function and the integrated correlation time for the energy. If we want to estimate the average position, we should do the similar data analysis but now for the position. This is quite tedious. Fortunately, there are simpler methods. Next, we discuss the blocking method and the method of independent runs.

Both the blocking method and the independent runs are clever ways of applying the central limit theorem. In the blocking method, the full data are divided into a number of equally sized blocks.
Here, we require that the block size is sufficiently large with respect to the correlation times, and also there is a good number of blocks like $30$-$100$. The idea is that the blocks are approximately independent if they are large, since the correlation essentially presents only at the interfaces of the blocks, then we can do error estimation using the CLT by treating the blocks as \textit{iid} objects.

Suppose we have a sufficiently large data set of an observable $\{x_i\}$ from the MCMC sampling, and we would like to estimate the sample mean $\mean{x}$. If we split the data set into $M$ equally sized blocks, we can estimate a sample mean for each block as $\mean{\tilde{x}}_i, i=1, 2, ..., M$. As the blocks are approximately independent, the estimates of the sample mean and the errorbar of the sample mean according to the CLT are:
\begin{align}
    \langle\tilde{x}\rangle &=\frac{\sum_i \langle\tilde{x}\rangle_i}{M}, \\
    \tilde{\sigma}_{\langle\tilde{x}\rangle} &=\frac{\tilde{\sigma}_{\langle\tilde{x}\rangle_i}}{\sqrt{M}}.
\end{align}

How to decide in practice if the block size $B$ is good, or equivalently how to choose the block number $M$ for a data set of size $R$? If we know the correlation times, we know how to choose a block size that is sufficiently large. If $\tau_{\mathrm{int}} \approx 270$, then $B=10000$ is likely sufficiently good, if $M$ is also reasonably large. When we do not know the correlation time, we can find a good one by trial and error. First, we try the block size $B=1$, we are here treating correlated data as \textit{iid} random variables, this will underestimate the errorbar. Remember that the effective number of samples is smaller than $R$ when the data are correlated. Next, we gradually increase the block size, and the errorbar will increase as well. If our data set is sufficiently large, then, there is a block size beyond which the estimated error is approximately \textbf{independent} of the block size or block number. However, the block size should not be too large, we also need a good number of blocks to make the CLT happy. Therefore, the block size should be smaller than approximately $R/30$ if we want to have at least $30$ blocks. If these conditions are met, the estimated errorbar is honest and we can select a block size in the relevant intermediate regime. 

We can also do \textbf{independent runs}, the idea is quite similar to the blocking method. Here, we are treating the entire runs as \textit{iid} objects, and we run the MCMC simulation many times to directly observe the fluctuation of the sample means. Suppose we have $M$ averages from $M$ independent runs, each run has $R$ data.
Compute the standard deviation of the sample means, that is the error estimation of the sample mean \textbf{at the level of data size $R$}. 

It is a good idea to average these sample means of independent runs, and the errorbar of the total sample mean is the standard deviation of the sample means divided by a factor of $\sqrt{M}$. This is exactly the CLT, and the sample mean and its error is actually \textbf{at the level of data size $MR$}. The pertinent equations are the same as those of the blocking method, so we shall not repeat them again here. If one feels like, each independent run is a block. There are also other variants of methods, but the idea is very similar, see, e.g., \cite{UPA}.

If the independent runs are sufficiently long, then combining $M$ runs of size $R$ should be very similar to a single long run of size $MR$. However, doing many small independent runs is not a good idea because MCMC has a systematic error from the initial guess of the state.
There is a very simple way to distinguish the \textbf{statistical error} and the \textbf{systematic error}, the former can be suppressed by averaging over independent runs but the latter cannot. The statistical error is purely from the finite sample size, if we average over sufficiently many independent runs, it can be arbitrarily suppressed. 
\textbf{Systematic error is bias} in nature, it cannot be removed by averaging over independent runs.

Consider the famous $\pi$ experiment by randomly throwing points on a square \cite{AA:comp}, if we throw only $10$ points, we get an estimate of $\pi$. It is not particularly accurate, but this is statistical error, there is no bias. If you ask some friends to do the same experiment, each friend gives you $10$ data, and you average the results, the estimate of $\pi$ can become arbitrarily accurate as more data are collected.

Now consider instead sampling the geometric distribution using the MCMC, and similarly we only record $10$ data in the run, the estimate of $\mean{n}$ is also not particularly good. If we similarly ask our friends to conduct the MCMC runs of length $10$ and we collect and average all of the results, we can similarly suppress the statistical error. However, this time we have a problem, our data are biased. Suppose the friends are all MC experts, so they all choose to start from the natural initial state $n=0$. After $10$ steps, there is no chance for, e.g., $n=50$ to appear in the sampling. Note that this is true no matter how many independent runs we collect. 

Systematic error, however, can be suppressed in time. Since the correlation decays exponentially in time, we expect a random walk will forget of its initial state exponentially, i.e., for the state to become a typical state of the distribution. In a MCMC simulation, it is best to \textbf{run for a while before any data collection} such that the initial state is forgotten. If this time we ask our friends to run $10^4$ steps before collecting the $10$ data, and now we average over these data, the result should be pretty good, i.e., the systematic error this time should be very small, the result is dominated by the statistical error. In a typical MCMC simulation, e.g., for sampling the geometric distribution, if the correlation time is about $300$, we can sample $1010000$ data and keep the final $1000000$. Note that the extra work is only $1\%$ of the total data collection, it is not a big deal.

When the correlation times are unknown, we can estimate the time scales for thermalization by looking at the cumulative average, i.e., we look at how a sample mean evolves with time and levels off. Then, we can choose a time scale and discard the data before the time where the sample mean approximately levels off. In addition, one can also start from two very different initial states and check how the cumulative means converge from different directions. 
However, it should be noted that this time scale is typically (much) larger than the correlation time, so one may have thrown away more data than necessary this way. Nevertheless, if one discards the data before this thermalization time, the systematic error should be very small.

In summary, the MCMC method is a sampling technique by doing random walks in a state space by requiring both ergodicity and global balance. It is frequently convenient to satisfy the global balance by restricting the detailed balance. We have practised designing MCMC random walk rules for quite a number of systems, and also studied how to analyze the correlated data generated by the Markov chain. The MCMC is the central algorithm of the Monte Carlo method, and remains the key part of the more advanced Monte Carlo methods.
Finally, we mention that the simple uniform sampling can be viewed as a special case of the MCMC sampling. In this setting, we propose a totally new state in each step, and because the new state and the present state are equally good, i.e., they have the same probability, we always accept the move.

\section{Conclusions and Outlooks}
In this work we introduced the MCMC method using elementary distributions. We started from two states using a model of population dynamics, then we discussed three states, followed by the stochastic matrix and its basic properties. The setup was subsequently generalized to many states of both discrete and continuous distributions. Finally, we presented the data analysis of correlated data using various methods.
We have systematically introduced all the central MCMC concepts in this rather simple setting, effectively removing essentially all the technical details of more realistic physical models in physics.

Our work can be followed by the regular introduction to the Ising model and classical particles. In the former one can randomly select a spin and propose to flip it, while in the latter one can randomly select a particle and propose to shift it. Then, one can study their higher dimensional generalizations. In our lectures, we discussed these materials before the data analysis part. After introducing the MCMC, one can switch to more advanced Monte Carlo methods, e.g., extended-ensemble methods \cite{Hukushima:PT,SA,Machta:PA}, cluster methods for $O(n)$ magnets \cite{Wolff,houdayer:01}, and various quantum Monte Carlo methods \cite{VMC,DMC,Mittal_2020,Binder:MC,NP:MC}.

\begin{acknowledgments}
This article is rewritten from the Markov Chain Monte Carlo part of our Monte Carlo lecture notes. We thank our students for their invaluable feedbacks and dedicated efforts towards understanding the Monte Carlo method throughout the entire course in the fall of $2021$. 
We gratefully acknowledge supports from the National Science Foundation of China under Grant No. 12004268, the Fundamental Research Funds for the Central Universities, China, and the Science Speciality Program of Sichuan University under Grant No. 2020SCUNL210.
\end{acknowledgments}

\bibliography{Refs}

\end{document}